\documentstyle[12pt]{article}
\begin{document}
\def\strut{\rule[-.5cm]{0cm}{1cm}}
\def\dspace{\baselineskip = .30in}

\title{
\begin{flushright}
{\large \bf IFUP-TH 60/94}
\end{flushright}
\vspace{1.5 cm}
{\large\bf Biased discrete symmetry and domain wall problem }}

\author{G. Dvali $^{a,b)}$,
Z. Tavartkiladze $^{b)}$ and J. Nanobashvili $^{b)}$
\\ \\
a){\small\it Dipartimento di Fisica, Universita di Pisa and INFN,}\\
{\small\it Sezione di Pisa, I-56126 Pisa, Italy}\\
{\small\it e-mail: dvali@ibmth.difi.unipi.it}\\
b){\small\it Institute of Physics, Georgian Academy of Sciences,}\\
 {\small\it 380077 Tbilisi, Georgia}\\}

\maketitle

\begin{abstract}
We reconsider a cosmological evolution of  domain walls produced by
spontaneous breaking of an approximate discrete symmetry. We show,
that   domain walls may never collapse
even if the standard bound on the vacuum energy asymmetry
is satisfied.
Instead of disappearing, these defects may form  stable
``bound states'' - double wall systems.
Possible stability of such a wall
 is a dynamical question and consequently restricts the
allowed range of parameters. In particular, in the
two Higgs doublet standard model with an anomalous $Z_2$ symmetry,
the above restriction suggests the mass of the pseudoscalar Higgs
(would be axion) being close to the mass of the scalar one.

\end{abstract}
\newpage

\dspace

{\bf 1.~~Introduction}
\bigskip

 Spontaneously broken discrete symmetries often play a fundamental role
in  particle physics models. Unfortunately, in the cosmological
context such theories exhibit serious problems [1] as they lead to the
formation of the topologically stable domain walls [2].

Enormous energy stored in these structures very soon comes to dominate the
universe, unless the scale of discrete symmetry breaking (or the effective
self coupling constant of the corresponding Higgs field) is very small
($<10^{-2} GeV$ or so) [1,2].

One famous solution for the domain wall problem is inflation [4].
However, normally inflation deals with scales as high as
say $M_G \sim 10^{16} GeV$ (the GUT scale).  In this situation,
it is clear, that the topological defects formed
below $M_G$, and in particular thous attributed to the electroweak phase
transition, will survive. Therefore, it seems that at least for
the weak scale domain walls the inflation may not be a good solution.

Alternatively,
cosmological troubles caused by the spontaneously broken discrete symmetry
can be avoided if this symmetry is approximate, meaning that it is also
explicitly broken by relatively small amount. From the naturalness point
of view this `solution' only makes sense if explicit breaking results
dynamically from some underlying physics e.g.
 from the anomaly [4]. Another  possibility [5] is that
gravity may not respect ungauged discrete (or continuous) symmetries and
explicit breaking can manifest itself through the higher dimensional Planck
scale $(M_p)$ suppressed operators in the scalar potential.

Whatever the source of the explicit violation is, below certain temperature
it creates energy difference between initialy degenerated
vacua. Consequently one vacuum  becomes energetically more
favorable and corresponding domains start to expand pushing the walls
away, whereas  unfavorable domains tend to collapse and disappear. The
necessary condition for the domain walls to disappear before dominating the
universe has the following form [6]

\begin{equation}
\epsilon >(\frac{\sigma}{M_p})^2
\label{cond0}
\end{equation}

where $\epsilon$ is the vacuum energy density difference and
$\sigma$ is the wall energy per unit surface. In the simplest case,
when the domain walls are formed by a single component real Higgs field
$\Phi$ with vacuum expectation value (VEV) $<\Phi>=\pm V$, the condition
(1)
can be enough to guarantee that walls collapse without any
trouble for cosmology. In the case of the real scalar field the
expectation value  $<\Phi>$ always vanishes across the wall and two
neighbor walls (bounding collapsing domain from the opposite sides)
sooner or later will annihilate each other. However, in the realistic
theories the Higgs fields usually are represented by complex scalars and
have several components. The elements of the discrete symmetry group
 can be mimicked by certain abelian (or even nonabelian)
phase transformation (which of course is not a part of continuous gauge
group, since we are not interested in walls bounded by strings [7]). This
circumstance brings a qualitatively new point in the wall structure
implying  that there is no topological reason for the absolute value of
the Higgs field $|\Phi|$ to vanish inside the wall. In other words,
whether this will happen becomes a matter of the parameter choice in the
scalar potential. For a wide range of these parameters the
energetically most favorable path in the field space, connecting two
degenerated vacuums, never includes the point $|\Phi|=0$.
Thus, the Higgs field never vanishes
inside the wall and we can define its ``winding number''
 through the defect.

In the present paper we show that in such cases the domain walls may never
 collapse even if discrete symmetry is explicitly broken and (1) is
 satisfied. This has to do with the fact, that the walls of the collapsing
domain can annihilate each other only if the Higgs
 field winds oppositely through this walls. Two walls with the same winding
 can never annihilate each other and disappear. Instead,
 they form classically stable "bound states"-double wall systems. Expectation
 value of the Higgs field returns to its initial value after traversing such
 composite wall.
If winding is
parametrised by phase $\arg\Phi=\theta$, we will have
$\Delta\theta=2\pi$ through the
wall.  As far as we know, previously the $2\pi$-walls have been
considered only in the context of the walls bounded by cosmic strings
 [8], which often appear in the axion cosmology. In such case initialy
 there are no walls, but axionic string formed due to a global
 $U(1)_{PQ}$-symmetry breaking. $2\pi$-walls are formed later on when
$U(1)_{PQ}$
 gets explicitly broken (and only if there is no discrete subgroup left [9]).
 Each $2\pi$-wall is bounded by string. Existence of the string network
 crucially determines the cosmological evolution of the walls and makes
 them to decay without trace.

In contrast $2\pi$-walls discussed in the present paper can be
 cosmologically problematic, since there is no string network behind of
 their existence.  They are formed as an bound states of would be topologically
stable domain walls when laters collapse due to an explicit breaking of
discrete symmetry.

\bigskip
{\bf 2.~~General mechanism}
\bigskip

In this section we will study mechanism of the $2\pi$-wall formation on a
 simple model including one $SU(3)-$ color triplet quark $u_{L,R}$
 transforming under discrete $Z_2$ symmetry:

\begin{equation}
u_{L} \rightarrow u_{L}, \quad u_{R} \rightarrow -u_{R},
\label{z2sym}
\end{equation}

This quark gets a mass from the following Yukawa coupling
\begin{equation}
g_u \Phi \bar u_{L} u_R + h.c
\label{coup}
\end{equation}

were $\Phi= \Phi_R + i\Phi_I =|\Phi| e^{i\theta}$ is a complex Higgs scalar
($\Phi \rightarrow - \Phi$ under $Z_2$), whose VEV breaks $Z_2$ spontaneously.

We choose a tree level scalar potential in the form:
\begin{equation}
v=-\frac{M^2}{2} |\Phi|^2 + \frac{h}{4}|\Phi |^4 - \frac{m^2}{4}(\Phi^2+h.c.)=
-\frac{M^2}{2}|\Phi |^2+ \frac{h}{4}|\Phi |^4-
\frac{m^2}{2} |\Phi|^2 \cos{2\theta}
\label{v}
\end{equation}

For definiteness we assume all parameters $m^2,M^2,h$ being real and
 positive.  The possible quartic term $\Phi^4$ which is allowed by all
 symmetries we have excluded for the simplicity of the
 analysis. Note that first two terms in (4) respect global
 $U(1)$-symmetry $\Phi \rightarrow e^{i\alpha}\Phi$ which is explicitly
 broken by third one down to $Z_2 \subset U(1)$ and $m$ is the mass of
 would be goldstone boson.
The absolute value of the Higgs VEV is given by
$|\Phi|=V=(\frac{M^2+m^2}{h})^{\frac{1}{2}}$. This vacuum is discretely
 degenerated, since the potential is minimized by any $\theta=\pi N$ (where
 $N$ is integer). This degeneracy results in the formation of the
 topologically stable domain walls. Topological constraint forces $\theta$
 to change by $|\Delta\theta|= \pi$ across the wall. However, this
 defects are not truly stable, because the $Z_2-$symmetry is anomalous [4].

The important point about the wall structure is whether the absolute value
 of the scalar field vanishes  inside the wall.
Certainly, this is the case if $m^2>M^2$. In this situation the domain
wall solution is given by kink (antikink)
\begin{equation}
 \Phi_R = \pm V tanh [xV(h/4)^{1/2}];
\label{cond}
\end{equation}
($x$ is a coordinate perpendicular to the wall)
whereas the pseudoscalar component $\Phi_I$ is identically zero, since
its effective $[mass]^2$ term $M_I^2= h\Phi_R^2  + m^2 - M^2$, which
is spatially dependent trough the $\Phi_R$ VEV, is positive everywhere.
In such a case,
 the wall network will evolve along the lines discussed in [4,5]. That
 is we expect that after $Z_2$ will be explicitly broken by anomaly,
walls can collapse without any cosmological harm (provided (1)  is
 satisfied).

  The significant deviation from this scenario (to be
 discussed below) will occur if $|\Phi|$ stays nonzero everywhere
 including the vicinity  of the domain wall.
In this case the energetically most favorable trajectory connecting two
 neighbor vacuums $\theta=\pi N$ and $\theta=\pi(N \pm 1)$, goes through the
 saddle point  $\Phi_I = \pm [(M^2-m^2)/h]^{1/2}$ and $\Phi_R =0$,
whereas the point $|\Phi |=0$
 is maximum. Of course, the condition $M^2>m^2$ is not yet sufficient to
conclude that pseudoscalar will get a nonzero VEV inside the wall, since
the gradient energy wants $\Phi_I$ to stay constant everywhere. So one has to
check the stability of the small perturbation about $\Phi_I =0$ in the
background of kink. For example this can be done through the analysis of
linearized equation for the pseudoscalar mode with the potential
$M_I^2$ on the existence of the bound state along the lines
discussed by Witten [10] for the bosonically superconducting string.
Such consideration shows that condensation of pseudoscalar inside the
wall  occurs at least for

\begin{equation}
m^2 < M^2 \sim M^2-m^2
\label{cond1}
\end{equation}

In such a case it costs a lot of energy ($\sim M^4$ per volume) for
 $|\Phi |$ to vanish somewhere in the space and in particular in
 the wall vicinity.
 To approximate the wall structure it is useful to put
 $|\Phi|=V=constant$  for a moment. The equation of motion
 for $\theta$ becomes then a well known sine-Gordon equation:

\begin{equation}
2 \frac{\partial^2\theta}{\partial\xi^2}
-\sin2\theta=0
\label{gordon}
\end{equation}
(where $\xi=mx$ and $x$ is the coordinate transverse to the wall)
 which has a topologically stable solution (soliton)

\begin{equation}
\theta(\xi)=2\cdot tan^{-1} exp(\xi).
\label{soliton}
\end{equation}

This is a reasonably good approximation for our purposes, although in reality
$|\Phi|$  can not stay constant across the wall due to a back
reaction.  In turn this will alter the shape of $\theta(\xi)$ , but whatever
 the exact form of $\theta(\xi)$ is, it should change by
 $|\Delta\theta|=\pi$ across the soliton.

At the QCD scale the color instanton effects become important
 and
 they explicitly break $Z_2$-symmetry. The effective $Z_2$-noninvariant term
 generated in the scalar potential by instanton vertex has the form

\begin{equation}
V_{inst}= - m_u\Lambda^3\cos\theta
\label{inst}
\end{equation}

where $\Lambda$ is a typical scale factor of the order the QCD scale
$\Lambda_{QCD} \sim 200 MeV$ and $m_u$ is the quark mass.

This term creates energy density difference $\epsilon=2m_u\Lambda^3$
 between $\theta=2N\pi$ and $\theta=(2N+1)\pi$ vacuums.
This energy difference creates pressure difference which drives the domain
 walls. Assuming say $m_u\Lambda^3 > 0$,
 the preference is given to $\theta=2N\pi$  domains which tend
 to expand, whereas thous with
 $\theta=(2N+1)\pi$  will tend to collapse and disappear. The dynamics of
the collapsing walls is defined by several factors. At early stages (when
explicitly violating effects are negligible) this are mainly wall tension
and the frictional force. Crudely the time dependences of this two factors
are proportional to $\sim (\sigma M_P^2/t^3)^{1/2}$ and $\sim (M_P/t)^2$
respectively. At some point the system becomes dominated by pressure
difference (with corresponding force per unit area $\sim \epsilon$)
which forces the false vacuum to collapse. For walls to  disappear,
above has to happen before they dominate the universe and we are lead to the
lower bound (1) for the vacuum energy asymmetry (for more details see
[2,4,5,6]).

 Due to above arguments in general one expects [4] the domain walls
disappearing soon after the QCD phase transition, since for the
weak scale walls (destabilized by QCD anomaly) condition (1) is satisfied
($\epsilon (M_P/\sigma)^2 \sim 10^{20}$ or so).

The similar
conclusion was made in [5] for the case of the domain walls destabilized
by "gravity induced" higher dimensional operators.
 However as it will be shown in the present paper, the resulting dynamics
of the walls is more sensitive to the parameters than one can naively
expect from (1). Namely for the wide rang of parameters (forming the
subrange of (1)) the walls may not collapse completely. In particular
this can happen for (6) meaning, that the mass of the pseudoscalar
mode (would be axion) is more than twice less the scalar one (radial mode).
In such a case only the walls having $opposite$ signs of $\Delta \theta$
can annihilate each other, whereas others can not. Instead the laters will be
paired forming stable  double wall systems
across which $\vert \Delta \theta \vert = 2\pi $. Such a wall pairing can
never occur if $\vert \Phi \vert$ vanishes across the
wall. In this  case the vacuas $\theta = 2\pi N$
(or $\theta = (2N+1)\pi$ ) corresponding to different $N$ are identical.
So in fact we have just two possible domains (say $(+)$ and $(-)$
respectively). This has to do with the fact that actually $\theta$ does
not "winds" through the wall, since it becomes ill defined at
$\vert \Phi \vert = 0$. The planar domain wall solution separating
$(-)$ and $(+)$ ($(+)$ and $(-)$) domains is given by kink (antikink) (5)
rather then by soliton (8).
 Now if we switch on the instanton effects the $(-)$ domains
will start to collapse and finally disappear, since the two neighbor
domain walls (kink and antikink) will annihilate each other.

Situation will drastically change if nonzero $\Phi$ is
energetically favored everywhere in the space.
That is, for example, if we are in the range of parameters given in (6).
Now $\theta$ is well defined for any spatial point $x$
 (including thous in the vicinity of the
domain wall) and as we show in the next section,
this fact plays a crucial role in the subsequent
evolution of the system.

\bigskip
{\bf 3.~~Cosmological evolution of $|\Phi| \neq 0$ domain walls}
\bigskip

 Now let us follow the history of the
early universe in our model. At high temperature ($T >> M, m$) the
$\Phi$-dependent part of the potential receives correction (Yukawa interaction
is neglected) [11]

\begin{equation}
 \Delta V(T) = |\Phi|^2 T^2h/6
\end{equation}

and we see that mass difference between the scalar and pseudoscalar fields
($m^2$) never vanishes. Therefore, at the moment of the phase transition,
when $\Phi_R$ and $\Phi_I$ pick up a nonzero VEVs, they already ``know''
about explicit breaking of global would be $U(1)$-symmetry. At this stage,
probability of string-bounded wall formation is exponentially suppressed
by the factor $mR$, where $R$ is a size of defect. Since the rang of our
interest is $m$ being not much (say within one order of magnitude)
below $M$, the probability of large string-wall formation is negligible.
So we are left with a network of domain walls of very complicated
geometry. Traversing each wall, $\theta$ winds by
$\Delta \theta = \pm \pi$ continuously interpolating between two values that
differ by $\pi$. Difference among "winding numbers" of neighbor walls
becomes very important after explicit breaking of the discrete symmetry.
Instanton induced term (9) in the scalar potential gives preference to a
certain domains (say to $\theta = 2\pi N$) which tend to expand. Crucial
point however is that not all $\theta = (2N+1)\pi$ domains will disappear.
To see this consider collapsing domain $\theta = \pi$ bounded by two infinite
planar walls. It is clear that this boundaries can  annihilate each other
and disappear   only if they correspond to the opposite windings of $\theta$.
That is two neighbor domains (of $\theta = \pi$ one) must have
both $\theta =0$ or both
$\theta = 2\pi$ (and never one $0$ and another $2\pi$). Contrastly, if
$\pi$-domain was initialy created between $0$ and $2\pi$ ones, the
corresponding walls can never annihilate each other since they have
both the same $\Delta \theta$. Instead of annihilation such walls will create
stable bound states. Double wall solution can
be approximated analytically if we assume $|\Phi| = constant$.
In such case equation for $\theta$ is double sine-Gordon equation.

\begin{equation}
2\frac{d^2\theta}{d\xi^2} -sin 2\theta -asin\theta=0
\end{equation}

where $a= {m_u\Lambda^3 \over V^2 m^2}$.
The composite wall can be approximated by stable double-soliton solution

\begin{equation}
\theta=2arccos \frac{a/2-Z^2}{2Z},~~~
Z^2 =2+a/2+2\sqrt{1+a/2}th(2\sqrt{(1+a/2)}\xi).
\end{equation}

Note that the usual sine-Gordon equation (7) also admits
 $2\pi$-soliton solution
 which is however unstable and tends to decay into two
solitons with $\Delta\theta = \pi$ each. The last term in (11) oposses this
decay and prevents two $\pi$-solitons from escaping each other,
forming the bound state. The width of the bound state (typical distance
between $\Delta\theta=\pi$ walls) is $\delta \simeq {1 \over m}ln{8\over a}$.
Since $\Lambda$ and $V(\sim m)$ correspond to QCD and weak scales respectively,
this width is by order of magnitude larger the width of the constituent
$\pi$-solitons $\delta ' \sim {1 \over m}$.
Infinite planar $2\pi$-wall
is not topologically stable and can decay quantum mechanically.
This decay goes through the tunneling process which creates a hole
 connecting $\theta = 0$ and $\theta = 2\pi$ vacuums with each other.
Imagine a closed path that starts in one of the domains (say in
$\theta = 0$), goes once through the hole and finally returns back to the
base point by traversing the wall in some other place.
 The one who travels along  the  path, will find phase $\theta$
changing by $\Delta \theta = 2 \pi$ at the end of the journey. Thus,
$|\Phi|$ has to vanish at some point inside the region encircled by
the path. By moving path around continuously (without
crossing the edge of the hole), the point $|\Phi | = 0$ will follow
a closed line. So, there
is a loop of the cosmic string enclosing the hole somewhere. Since, the
space between two solitons is initialy filled with $\theta = \pi$ phase,
the formation of the hole is costly in the energy of its own edge. The
later includes the  energy of the (cylindric) wall surface that
bounds the hole and the energy of the string loop that is somewhere in
this boundary. Hole will start to expand classically when, this energy will
be overcomed by energy of the  punched out piece of domain wall.
The change in the energy caused by $R$-radius hole formation can be
 estimated to be:

\begin{equation}
  \Delta E = \pi R [2(\mu^2 + \delta \sigma) - R(\delta \epsilon +
2\sigma)]
\end{equation}

where $\mu^2 \simeq 2\pi V^2ln{M \over m}$ and $\sigma \simeq 4mV^2$ are
string and wall tensions respectively. Hole expands when its
radius exceeds the critical value
 $R_c= (\delta + \mu^2/\sigma)/2$.  The tunneling rate is
exponentially suppressed by the factor
$2\pi R_c^3 \sigma \sim \pi [ln{8 \over a} ({M \over m})^{\pi /2}]^3 (V/m)^2$.
So that even for $V/m \sim 1-10$ this wall can be stable
for all practical purposes.  Such $2\pi$-wall of the horizon
size (if formed) can be disastrous cosmologically.
To destabilize this structures we should increase parameter $m^2$ and convert
$\vert \Phi\vert = 0$ in to the saddle point. In the other  words we have to
increase the mass of angular pseudoscalar mode (would be goldstone boson)
relative to scalar one (radial mode).
Closer $\vert \Phi\vert = 0$ is to the saddle point, less
costly is the string bounded hole formation in the wall sheet. In the limit
when $\vert \Phi \vert = 0$ becomes a saddle point, $2\pi$-walls become
classically unstable.

\bigskip
{\bf 4.~~ Two doublet extention}
\bigskip

Above model is trivially extendable to the realistic example of ref [4].
This is
the $SU(2)\otimes U(1)$ standard model with two Higgs doublets $\Phi_u ,\Phi_d$
and anomalous discrete symmetry $Z_2$ needed for natural flavor conservation.

Under $Z_2$:
\begin{equation}
\Phi_u \rightarrow -\Phi_u, \quad \Phi_d \rightarrow \Phi_d, \quad
Q_L \rightarrow Q_L, \quad U_R \rightarrow -U_R, \quad
d_R \rightarrow d_R
\end{equation}

where $Q_L =\left(\matrix{u\cr
                          d\cr}\right)$ is the left-handed quark doublet.
 $Z_2$-symmetry ensures that each Higgs
doublet couples only to the fermions of the given charge.
The scalar potential now takes the form

\begin{equation}
V=\sum_{\alpha}(M_{\alpha}^2 |\Phi_{\alpha}|^2 +
h_{\alpha}|\Phi_{\alpha}|^4) -
h|\Phi^*_u \Phi_d|^2 - h'|\Phi_u|^2|\Phi_d|^2+ V_{phase}
\end{equation}
\begin{equation}
V_{phase} = -{\lambda \over 2}(\Phi^*_u \Phi_d )^2 + h.c
\end{equation}

where $\alpha = u,d$.
If $h>0$ the VEVs are pointing in the electrically neutral direction
$\Phi_{\alpha} = \left(\matrix{ V_{\alpha} e^{i\theta_{\alpha}}\cr
                         0\cr }\right)$.
In this parameterization

\begin{equation}
V_{phase} = -\lambda V^2_u V^2_d cos(2\theta) \quad
\theta = \theta_u- \theta_d
\end{equation}

For $\lambda>0$, $V_{phase}$ is minimized by $\theta = \pi k$ $(k=1,2...)$
and there are topologically stable domain walls. If  $h+h'$ is
 larger than $\lambda$, nonzero expectation values
$V_u,V_d \neq 0$ will be energetically
flavored everywhere in the space and even in the
 vicinity of the domain wall. So that as before, the winding of
$\theta$ can be well defined.
Explicit violation of $Z_2$-symmetry by QCD instantones leads to the
formation of "composite" $2\pi$-walls much in the way discussed in Sec.2.
The source of energy difference induced by instanton vertex in scalar
potential has the form (for simplicity we assume just one family of quarks):

\begin{equation}
V_{phase} = - m_u m_d \Lambda^2 cos \theta
\end{equation}

where as before $\Lambda\sim\Lambda_{QCD} \simeq 200MeV$ and $m_u, m_d$
are masses of $u$ and $d$-quarks respectively. The resulting
double sine-Gordon equation for $\theta$ has the form (11), where now
$a= {m_u m_d \Lambda^2 \over V^2 m^2}$.
Once again, the resulting $2\pi$-walls can be practically stable and
cause trouble for the cosmology, unless the parameter $\lambda$ is
sufficiently large implying, that the mass of the  would be axion
is close to the Higgs mass.

\bigskip
{\bf 4.~~Conclusions}
\bigskip

 We have shown that for the certain rang of parameters, compatible
with standard bound, the collapse of  unstable domain walls may end
up with formation of classically (and practically) stable double wall
systems. If formed, this structures can be problematic cosmologically
and therefore their nonexistence requires additional restriction of the
parameter space. Of course, formation and subsequent evolution of this
defects is very complicated process and requires much more detailed
and perhaps numerical study. For example, collision of the domain walls
may enhance the rate of hole formation.
At this point our analysis can be considered
as `first' approximation, but nevertheless, it provides an interesting
enough example how cosmological considerations may restrict parameters
in the particle physics models.

\bigskip
{\bf Acknowledgment}
\bigskip

One of us (G.D.) would like to thank Goran Senjanovic and Vazha Berezhiani
for discussions and INFN for support.

\newpage
\centerline{\bf References}
\begin{description}
\item{[1]} Ya.B.Zel'dovich, I.Yu.Kobzarev and L.B.Okun,
JETP 40 (1975) 1.
\item{[2]} For a review, see: T.W.B.Kibble, J.Phys. A9 (1976) 1387;
Phys.Rep. 67 (1980)  183; A.Vilenkin, Phys.Rep. 121 (1985) 263.
\item{[3]} A.H.Guth, Phys.Rev.D23 (1981) 347;
 For a review, see e.g. A.D.Linde, Particle Physics and Inflationary Cosmology
 (Harwood Academic, Switzerland, 1990); K.A.Olive, Phys.Rep. 190 (1990) 307.
\item{[4]} J.Preskill, S.P.Trivedi, F.Wilczek and M.B.Wise,
 Nucl.Phys.B363 (1991) 207.
\item{[5]} B.Rai and G.Senjanovic, Phys.Rev. D49 (1994) 2729; For earlier work
           on the possible role of gravity, see
B.Holdom, Phys.Rev. D28 (1983) 1419
\item{[6]} A.Vilenkin, Phys.Rev.D23 (1981) 852.
\item{[7]} T.W.B.Kibble, G.Lazarides and Q.Shafi, Phys.Rev D26 (1982) 435.
\item{[8]} A.Vilenkin and A.E.Everett, Phys.Rev.Lett. 48 (1982) 1867;
         For a review, see e.g. J.E.Kim, Phys.Rep. 150 (1987) 1.
\item{[9]} P.Sikivie, Phys.Rev.Lett 48 (1982) 1156.
\item{[10]} E.Witten, Nucl.Phys. B249 (1985) 557.
\item{[11]} S.Weinberg, Phys.Rev. D9 (1974) 3357;
            L.Dolan and R.Jackiw, Phys.Rev. D9 (1974) 3320;
            A.D.Linde, Rep.Prog.Phys. 42 (1979) 389.
\end{description}
\end{document}